\documentclass[11pt]{article}

\usepackage{amsfonts}%

\begin{document}

\newcommand {\ignore} [1] {}

\newtheorem{theorem}{Theorem}[section]
\newtheorem{lemma}[theorem]{Lemma}
\newtheorem{fact}[theorem]{Fact}
\newtheorem{corollary}[theorem]{Corollary}
\newtheorem{definition}{Definition}[section]
\newtheorem{proposition}[theorem]{Proposition}
\newtheorem{observation}[theorem]{Observation}
\newtheorem{claim}[theorem]{Claim}
\newenvironment{proof}{\noindent{\bf Proof:\/}}{\hfill $\Box$\vskip 0.1in}
\newenvironment{proofsp}{\noindent{\bf Proof}}{\hfill $\Box$\vskip 0.1in}

\pagenumbering{arabic}

\date{}

\title{Influence is a Matter of Degree: New Algorithms for Activation Problems \thanks{Work supported in part by The Israel Science Foundation (grant No. 873/08).}}

\author{ \\Daniel Reichman
\thanks{Weizmann Institute of Science, PO Box 26, Rehovot 76100,
Israel. Email: \texttt{daniel.reichman@gmail.com}}
 }

\maketitle

\def\opt   {\mathsf{opt}}

\begin{abstract}
We consider the target set selection problem. In this problem, a vertex is active either if it belongs to a set of initially activated vertices or if at some point it has at least $k$ active neighbors ($k$ is identical for all vertices of the graph). Our goal is to find a set of minimum size whose activation will result with the entire graph being activated. Call such a set \emph{contagious}. We prove that if $G=(V,E)$ is an undirected graph, the size of a contagious set is bounded by $\sum_{v\in V}{\min \{1,\frac{k}{d(v)+1}\}}$ (where $d(v)$ is the degree of $v$). We present a simple and efficient algorithm that finds a contagious set that is not larger than the aforementioned bound and discuss algorithmic applications of this algorithm to finding contagious sets in dense graphs.
\end{abstract}

% \noindent
% {\em Key words:} Wireless networks, Power assignment, Graph-connectivity, Approximation algorithms.

\section{Introduction} \label{s:intro}

The study of diffusion processes of ideas, products and trends within social networks is the subject of intense research \cite{Jack}. With the growing influence of the WWW on social interactions and the increasing popularity of blogs as well as sites such as Facebook and Linkedin, it is widely accepted that dynamics of information spread have crucial influence on social and economical phenomena \cite{Klei07,Morris}. Moreover, understanding the nature of such dynamics has great practical significance as it can increase the effectiveness of viral marketing methods where one seeks to find influential individuals within a network realizing that the adoption of products by such individuals can have highly beneficial effects to a marketing campaign \cite{JLM01a,JLM01b}.

One frequent way to model diffusion processes is threshold models \cite{Gra,KKT03}. In these models we have a directed or undirected graph $G=(V,E)$, where nodes represent individuals and edges represent relationships between individuals within the network. We assume that a specified subset of vertices is initially activated, and that each vertex has activation function which depends on the number of active neighbors. Once the fraction of activated neighbors of an inactive vertex reaches a certain threshold, the vertex becomes activated. The processes is progressive-when a vertex is activated it stays active forever. We will sometime say "infected" rather than activated, and these two terms will be used interchangeably throughout this paper.

Perhaps the simplest threshold models are those in which inactive vertex $u$ becomes active iff a fixed number of $u$'s neighbors, say $k$, become active. In this work, we assume that $k$ is identical for every vertex. $k$ may depend on the number of vertices of the graph, although more attention has been devoted to the case where $k$ is a small constant such as $2$ or $3$.

A natural question, given an undirected graph and a parameter $k$, is how many vertices need to be activated in order to "infect" the entire set of vertices. Such questions were studied in probability theory and combinatorics: boostrap percolation problems are concerned with the minimal probability $p$ such that if every vertex is independently activated with probability $p$, the entire graph is activated with high probability. Such problems are generally difficult and the exact answer is only known for certain special families of graphs such as grids \cite{BP}, hypercubes \cite{BBM}, random regular graphs \cite{BPit} and random graphs with a given degree sequence \cite{Jan}. A related problem is that of activating the smallest set such that the entire graph is activated. We refer to this optimization problem as Target Set Selection \cite{Chen09}. Unfortunately, Target Set Selection is NP-hard. Furthermore, the problem is hard to approximate within factor of $2^{\log^{1-\epsilon}n}$, where $\epsilon$ is arbitrarily small positive constant, even when the threshold of each vertex is two, and the degree of the graph is bounded by a constant \cite{Chen09}. To the best of our knowledge, no approximation algorithm achieving approximation ratio significantly smaller than $|V|$ is known for this problem. Finally, spread maximization problems, where one seeks to activate a set of cardinality $l$ that will infect the largest number of vertices have been also studied \cite{ES07,KKT03,KKT06}.

\section{Our Contribution}
One approach in dealing with hard problems on graphs is to consider properties of their graphical sequences (the sequence of degrees of vertices) as well as other properties of the degrees such as the maximal and minimal degrees. Oftentimes one can find interesting dependencies between the degree sequence, the minimal degree (or maximal degree) and bounds attained by approximation algorithms. For example, a simple randomized algorithm can find an independent set of size $\sum_{v\in V}{\frac{1}{d(v)+1}}$ in any undirected graph \cite{AS, Gr}, and a well known probabilistic algorithm finds a dominating set of size $O(\frac{\ln d}{d}n)$ in an undirected graph with $n$ vertices and minimal degree $d$ \cite{AS}.

In this work we show that similar approaches work for the target set selection problem. We first show that if $G$ has minimal degree $d$, all vertices can be infected by activating a set of size $O(\frac{n}{d})$ provided that $k=2$. We then proceed and apply the probabilistic method in proving a generalization of the the above bound:
\newline
 \begin{theorem} \label{t:main}
 Let $G=(V,E)$ an undirected graph with vertex set $\{v_1,...,v_n\}$. Denote the degree of the $ith$ vertex by $d(v_i)$, and assume that a vertex is activated iff it belongs to an initially infected set or at least $k$ of his neighbors are activated during the activation process. Then, one can activate the entire graph by activating at most $\sum_{v\in V}{\min\{1,\frac{k}{d(v)+1}\}}$ vertices. Furthermore, one can find such a set in deterministic polynomial time.
\end{theorem}

It is easy to see that this bound is tight (consider $\frac{n}{l}$ disjoint cliques of size $l$). While our proof of existence in theorem ~\ref{t:main} is based on a simple probabilistic argument, our algorithm is based by an algorithm implicit in \cite{AKS}. Finally note, that theorem ~\ref{t:main} implies that if the minimal degree of $G=(V,E)$ is $\Omega(|V|)$, then target set selection can be solved optimally in polynomial time. To the best of our knowledge, it was unknown whether this problem admits a polynomial time algorithm when the graph is very dense (e.g., has minimal degree which linear in the number of vertices).

\section{Preliminaries}

All graphs in this work are undirected with vertex set $\{v_1,...,v_n\}$. The \emph{degree} of a vertex is denoted by $d(v)$. A $k$\emph{-dominating set} is a set $D$ such that all vertices of $G$ are either in $D$ or have at least $k$ neighbors in $D$ (Thus a $1$-dominating set is simply a dominating set). A graph  $G$ is $d$\emph{-degenerate} if in every induced subgraph of $G$ there is a vertex of degree smaller than $d$.

Let $G=(V,E)$ be an undirected graph with $n$ vertices and let $k$ be a natural number that may depend on $n$. An infectious process is defined as follows: an initial set of vertices is activated. The processes progresses on discrete rounds. In each round, every inactivate vertex is activated iff at least $k$ of his neighbors are activated. $k$ is called the \emph{threshold} of $G$.  An activated vertex stays active throughout this process. We call a set $A \subseteq V$ \emph{contagious} if activating all vertices within $A$ results in activation of the entire graph. Finally we are interested in the following optimization problem:

\noindent
{\sf Target Set Selection} \\
{\em Instance:} \
An undirected graph $G=(V,E)$, with threshold $k$. \\
{\em Objective:}
Find a minimum cardinality contagious set.

\section{Warmup: the case of $k=2$}

In this section we show that given an undirected graph $G=(V,E)$ ($|V|=n$) of minimal degree $d$, if $k=2$ there is always a contagious set of size $O(\frac{n}{d})$. We assume that $d$ is "large enough" (note that if $d$ is a small constant, there is trivially a contagious set that is not larger than the above bound). Although the bounds in this proof are not as general as our main theorem, we present it here for two reasons: first it gives an intuitive explanation as to why activating $O(\frac{n}{d})$ vertices is sufficient to activate the entire graph. Second, the proof idea is different from the main theorem, and may prove to be useful in other contexts.

Notice that the activation problem for $k=2$ is similar to the $2-$dominating set problem: indeed a $2$-dominating set is always contagious. It is well known that every graph contains a $2$-dominating set of size $O(\frac{\ln d}{d}n)$: \cite{HaHenn,GaZve} if we activate every vertex independently with probability $p$, the probability a given vertex is not covered by two vertices is at most $(1-p)^{d+1}$ +$pd(1-p)^d$ which is approximately $e^{-p(d+1)}+pde^{-pd}$. It follows that the expected size of uninfected vertices is at most $n(e^{-p(d+1)}+pde^{-pd})$. Hence if $p=\frac{\ln (d+1)}{d+1}$ we have that the adding each uninfected vertex to the the set of activated vertices, results with a $2-$dominating set whose expected size is $O(\frac{\ln d}{d}n)$.

It turns out that we can save a $\ln d$ factor in the above bound by the following argument. Activate each vertex with probability $p$, but set $p$ to equal $\frac{1}{d}$. As before, the probability of a vertex not being activated is approximately $e^{-p(d+1)}+pde^{-pd}=e^{-1}(1+e^{-1/d})$ which is very close to $\frac{2}{e}$. Hence the expected size of the set the vertices that remain uninfected (denote it by $I$) is $\frac{2}{e}n$. By definition of the activation rule, every vertex in $I$ is connected to at least $d-1$ vertices in $I$. Hence we can apply the same argument on $I$ where the minimal degree is $d-1$ instead of $d$. Thus, by activating additional number of approximately $\frac{2n}{e(d-1)}$ vertices, the expected size of uninfected vertices in the second phase, $I'$ will decrease to $(\frac{2}{e})^2n$. Like the first phase, the minimal degree of each vertex in the subgraph induced by $I'$ would be at least $d-1$. Continuing in this fashion results with activating at most $\frac{n}{d-1}(\sum_{i=0}^{\infty}{(\frac{2}{e})^i})=\frac{n}{d-1}\frac{e}{e-2}$ vertices which consist of a contagious set whose size is $O(\frac{n}{d})$. As required

\section{Proof of the Main Theorem}

Consider a permutation $\sigma$ chosen uniformly at random from the set of all permeations of the vertices of an undirected $G=(V,E)$ with $V=\{v_1,...,v_n\}$. Let $L_i$ ($1 \leq i \leq k$) be the set of all vertices that appear in the $ith$ location among themselves and their neighbors, where the order between vertices is determined by the order in which they appear in the random permutation. For $v \in V$, $v$ belongs to $L_i$ with probability $\frac{1}{d(v)+1}$ if $i \leq d(v)+1$ and $0$ otherwise. Hence, $v\in \bigcup_{i=1}^k{L_i}$ with probability $\min (1,\frac{k}{d(v)+1})$. By the linearity of expectation, the expected size of $L=\bigcup_{i=1}^k{L_i}$ is $\sum_{v\in V}{\min\{1,\frac{k}{d(v)+1}\}}$. Activate all vertices in $L$. Every vertex either belongs to  $L$, or has at least $k$ neighbors that are eventually activated when it is activated as well. Hence by standard expectations arguments we have established the existence of contagious set whose size is at most $\sum_{v\in V}{\min\{1,\frac{k}{d(v)+1}\}}$

Before proceeding some simple observations are in order. First, it is easy to see that $L$ is in fact a $k$-degenerate graph. To see this, assume that the permutation of the vertices is $u_1,...,u_n$ where $u_i=v_{\pi^{-1}(i)}$. Then, if $\{u_{i{_1}},...,u_{i_l}\}\subseteq L$ with $i_1<i_2<...<i_l$, $u_{i_l}$ has degree smaller than $k$ in the subgraph induced by $\{u_{i{_1}},...,u_{i_l}\}$, proving the claim. Hence, we always have a contagious set that is in fact $k$-\emph{degenerate}. Moreover, the aforementioned proof implies that every undirected graph has a $k$-degenerate subgraph of size at least $\sum_{v\in V}{\min\{1,\frac{k}{d(v)+1}\}}$. This was proved by Alon, Khan and Seymour (e.g., \cite{AKS}). Our proof is different and somewhat simpler from theirs.

Finally we prove that one can find such a set deterministically.
Consider the following algorithm: While there is a vertex of degree at least $k$, choose among all vertices of degree at least $k$ the one with \emph{minimal} degree and delete it. When all vertices are of degree strictly smaller than $k$, stop and return these vertices as the set to be activated. Note that this algorithm runs in polynomial time. Denote the set of vertices this algorithm activates by $I$.

Clearly $I$ is contagious (the vertices not in $I$ are activated in reverse order to the order in which they were deleted).

It remains to prove that $|I|\leq \sum_{v\in V}{\min\{1,\frac{k}{d(v)+1}\}}$. We apply similar ideas to those used in \cite{AKS,Gr}. For an undirected graph, $G=(V,E),$ denote $ \sum_{v\in V}{\min\{1,\frac{k}{d(v)+1}\}}$ by $w(G)$. We show that $w(G)$ will not increase as vertices are deleted from $G$ during the iterations of our algorithm. When the algorithm terminates, $w(G(I))=|I|$ (where $G(I)$ is the graph induced by the vertices of $I$, the set activated by our algorithm). Hence $|I| \leq w(G)$, As required.

If all vertices are of degree at most $k-1$ then $|I|=w(G)$ and we are done. Otherwise, let $u$ be a vertex that is deleted in the first iteration, with $d(u)=\delta$. Denote $u$'s neighbors with degree at least $k$ by $\{u_1,...,u_l\}$ (this set may be empty). Since the contribution of the neighbors of $u$ with degrees at most $k-1$ is identical in both $w(G)$ and $w(G \setminus \{u\})$,  we have: \newline
$w(G \setminus \{u\}) = w(G)-\frac{k}{\delta+1}+\sum_{i=1}^{l}({\frac{k}{d_G(u_i)}-\frac{k}{d_G(u_i)+1}})  =w(G)-\frac{k}{\delta+1}+\sum_{i=1}^{l}{\frac{k}{d_G(u_i)(d_G(u_i+1))}} \leq w(G) -\frac{k}{\delta+1}+\frac{\delta k}{\delta(\delta+1)}= w(G)$. A simple inductive argument shows that $w(G)$ does not increase through the execution of this algorithm, concluding the proof.

\vspace{0.1cm}

\noindent \textbf{Conclusions}
\vspace{0.1cm}

We have proven a useful bound on the size of a contagious set in an undirected graph. These bounds implies that on dense instances one can find optimal (or near optimal) solutions to the Target-Set-Selection problem in polynomial-time. Unfortunately, in graphs whose degree is distributed according to power-law distributions that frequently occurs in real life networks \cite{Jack}, it is unlikely that all nodes will have very high degrees. An obvious direction for future research is to devise approximation algorithms for this problem where the graph is sparse. In addition, although our work implies a polynomial time algorithm for finding a contagious set in dense graphs, the running time of a naive algorithm that examines all subsets that are no larger than our bound and checks whether they are contagious, is rather large: even if we only go over all subsets of size $2$, the running time is $O(n^4)$. Finding more efficient algorithms that detect contagious sets of constant size is an interesting direction for future research.
\newpage

\vspace{0.1cm}
\noindent \textbf{Acknowledgements}
\vspace{0.1cm}

I would like to thank Shiri Chechik, Uri Feige, Elchanan Mossel and Inbal Talgam for useful discussions.

{\small
\bibliographystyle{abbrv}
\bibliography{references}
}

\end{document}